\documentclass{aa}

\usepackage{graphicx}
\usepackage{txfonts}
\begin{document}

   \title{H$_2$/HD molecular data for analysis of quasar spectra in search of varying constants
	     \thanks{Tables 1, 2 and 3 are only available in electronic form
		     at the CDS via anonymous ftp to cdsarc.u-strasbg.fr (130.79.128.5)
		     or via http://vizier.u-strasbg.fr/viz-bin/VizieR?-source=J/A+A/622/A127}
	 }
   %\titlerunning{H$_2$/HD molecular data for analysis of quasar spectra}

   \author{W. Ubachs
          \inst{1}
          \and
          E. J. Salumbides\inst{1}
          \and
          M. T. Murphy\inst{2}
          \and
          H. Abgrall\inst{3}
          \and
          E. Roueff\inst{3}
          }

   \institute{Department of Physics and Astronomy, Vrije Universiteit, De Boelelaan 1081, 1081 HV Amsterdam, The Netherlands\\
            \email{w.m.g.ubachs@vu.nl}
         \and
             Centre for Astrophysics and Supercomputing, Swinburne University of Technology, Melbourne,
             Victoria 3122, Australia
         \and
            Sorbonne Universit\'{e}, Observatoire de Paris, Universit\'{e} PSL, CNRS, LERMA, 92190 Meudon, France
             }

   \date{Received 5 December 2018 / Accepted 24 December 2018}

% \abstract{}{}{}{}{}
% 5 {} token are mandatory

  \abstract
  % context heading (optional)
  % {} leave it empty if necessary
   {Absorption lines of H$_2$ and HD molecules observed at high redshift in the line of sight towards
   quasars are a test ground to search for variation of the proton-to-electron mass ratio $\mu$. For this purpose, results from astronomical observations are compared with a compilation of molecular data of the highest accuracy, obtained in laboratory studies as well as in first-principles calculations. }
  % aims heading (mandatory)
   {A comprehensive line list is compiled for H$_2$ and HD absorption lines in the Lyman
   ($B^1\Sigma_u^+$ - $X^1\Sigma_g^+$) and Werner ($C^1\Pi_u$ - $X^1\Sigma_g^+$) band systems up to the Lyman cutoff at 912 \AA.
   Molecular parameters listed for each line $i$ are the transition wavelength $\lambda_i$,
   the line oscillator strength $f_i$,
   the radiative damping parameter of the excited state $\Gamma_i$, and the sensitivity coefficient
   $K_i$ for a variation of the proton-to-electron mass ratio.}
  % methods heading (mandatory)
   {The transition wavelengths $\lambda_i$ for the H$_2$ and HD molecules are determined by a variety of advanced high-precision spectroscopic experiments involving narrowband vacuum ultraviolet lasers,
   Fourier-transform spectrometers, and synchrotron radiation sources.
   Results for the line oscillator strengths $f_i$, damping parameters $\Gamma_i$ , and sensitivity coefficients $K_i$ are obtained in theoretical quantum chemical calculations.}
  % results heading (mandatory)
   {A new list of molecular data is compiled for future analyses of cold clouds of hydrogen absorbers, specifically for studies of $\mu$-variation from quasar data.  The list is applied in a refit of quasar absorption spectra of B0642--5038 and J1237+0647 yielding constraints on a variation of the proton-to-electron mass ratio $\Delta\mu/\mu$ consistent with previous analyses.}
  % conclusions heading (optional), leave it empty if necessary
   {}


   \maketitle
%
%-------------------------------------------------------------------

\section{Introduction}

  After~\citet{Thompson1975} had pointed out that possible variations of the proton-to-electron mass ratio $\mu=m_p/m_e$ could be deduced from the spectrum of molecular hydrogen, \citet{Varshalovich1993} performed an early analysis setting a constraint on $\Delta\mu/\mu$ over cosmological time scales using a quasar absorption spectrum. The rationale of such an analysis is based on the fact that multiple H$_2$ absorption lines exhibit a different dependence on a possible variation of the fundamental constant $\mu$.
  While initial work was based on classical spectroscopic studies of  the hydrogen molecule~\citep{Abgrall1993a,Abgrall1993b}, the laser-based studies of H$_2$~\citep{Philip2003,Ubachs2004,Reinhold2006} thereafter increased the precision of such constraints, as did the improved quality of astronomical observations.
Several groups have been involved in analyzing H$_2$ in the line of sight of quasars searching for a possible $\mu$-variation~\citep{Ivanchik2005,King2008,Thompson2009,Malec2010,King2011,Wendt2011,Rahmani2013,Bagdonaite2014,Albornoz2014,Dapra2015}.
  The status of searches on a drifting fundamental constant $\mu$ has been reviewed~\citep{Ubachs2016,Ubachs2018}, combining analyses of ten quasar absorption systems leading to an overall constraint: the proton electron mass ratio has changed by less than 5 ppm (at 3 $\sigma$) for redshifts in the range $z=2.0-4.2$, corresponding to look-back times of 10-12.4 billion years.

  The analyses of astronomical H$_2$ spectra observed in the line-of-sight toward quasars and the extraction of a possible $\mu$-variation depend on the availability of molecular data. Such data have been published on various occasions, such as in the supplementary material to the paper by \citet{Malec2010}. However, we have found some inconsistencies and small errors in the data set, in particular for values of line oscillator strengths, and not all data were specified to the best accuracies now available. The purpose of this paper is to provide a comprehensive and accurate data set of H$_2$ and HD line parameters, to be used reliably in future analyses of $\mu$-variation from hydrogen absorption in the early universe.

\section{Molecular data for H$_2$}

In the analysis of astronomical spectra in a search for a varying $\mu,$ the H$_2$ and HD absorption lines are represented by four characteristic values, $\lambda_i$, $f_i$, $\Gamma_i$ and $K_i$. Availability of accurate values for the wavelengths $\lambda_i$ of the absorption lines is most crucial. Laboratory wavelengths were determined with the use of a narrow-band tunable laser in the vacuum ultraviolet range in molecular beam spectroscopic studies~\citep{Philip2003,Ubachs2004,Reinhold2006}. Thereafter, accurate level energies of the $B^1\Sigma_u^+$ and $C^1\Pi_u$ excited states in the H$_2$ molecule were measured via a stepwise method using Doppler-free laser techniques and Fourier-transform emission~\citep{Salumbides2008}. From these determinations of level energies, listed by ~\cite{Bailly2010}, the transition wavelengths of individual H$_2$ absorption lines in the Lyman and Werner bands were derived with high accuracy by combining with ground-state level energies \citep{Jennings1983}.
An improvement on the accuracy of ground-state level energies has been produced via \emph{ab initio} calculations including relativistic and quantum-electrodynamic effects~\citep{Komasa2011}. The latter calculations were experimentally tested in precision laser studies \citep{Salumbides2011,Dickenson2013} and found to be accurate to $10^{-4}$ cm$^{-1}$.
Here we present a new determination of combination differences between excited state level energies \citep{Bailly2010} and the ground-state level energies \citep{Komasa2011}, resulting in improved values for transition wavelengths of Lyman and Werner lines at accuracies of $\Delta\lambda/\lambda <3 \times 10^{-8}$. These accurate data for the wavelengths are labeled with "1" in Tables 1 and 2.
In view of the fitting and calibration accuracies obtained with the spectrographs mounted on the largest telescopes, these values for the laboratory wavelengths may be considered exact for the purpose of comparison with astronomical data; see also~\cite{Ubachs2016}.

The excited-state level energies of~\cite{Bailly2010} have some limitations, in that not all high-$J$ levels were probed up to $J=7$, and that the $C^1\Pi_u$ ($v=4,5$) levels are missing; here $J$ refers to the rotational angular momentum of the ground state. In such cases, data are adopted from the XUV-laser experiments \citep{Philip2003}, while information for P-lines was also extended to wavelengths for R-lines calculated via combination differences based on known ground-state levels~\citep{Ubachs2007}. Based on the improved ground-state energies of \cite{Komasa2011}, the latter determinations have been verified and improved for accuracy, estimated at $\Delta\lambda/\lambda \sim 10^{-7}$. These entries for the wavelengths in Tables 1 and 2 are labeled with "2". Also uncertainties of the wavelengths are listed for all entries.

In some remaining cases, the wavelengths were taken from classical spectroscopic studies by Abgrall et al. (1993c). In these cases the accuracy is lower and estimated at $\Delta\lambda/\lambda \sim 10^{-6} - 10^{-5}$. The latter entries are labeled with "3" in Tables 1 and 2.

Line oscillator strengths $f_{v''J'',v'J'}$ involve the square of the transition dipole moment $M^{\alpha}$ matrix elements connecting the ground $X^1 \Sigma_g^+$ ($v'',J''$) and the ($v',J')$ electronic excited states as well as the transition energy:
\begin{equation}
f_{v''J'',v'J'}= \frac{2}{3(2J''+1)}(E_{v'J'}-E_{v''J''}) |M^{\alpha}|^2
.\end{equation}
All terms  are expressed in atomic units, $f$ is dimensionless, and $\alpha$ corresponds to P, Q, and R transitions where $\Delta J = J' - J'' = -1, 0 ,+1,$ respectively. The matrix elements of the transition moments are computed with the wave functions numerically calculated from the integration of the coupled Schr\"{o}dinger equations linking the $X^1\Sigma_g^+$ and $B^1\Sigma_u^+$, $C^1\Pi_u$, $B'^1\Sigma_u^+$, and $D^1\Pi_u$ excited electronic states as described in \cite{Abgrall1993c}. The computations include nonadiabatic rotational and radial couplings taking place between the four excited electronic states.
Einstein $A$ emission probabilities may be derived from the following expression (in CGS units):
\begin{equation}
  A_{v'J',v''J''}= \frac{4}{3\hbar^4c^3(2J'+1)}(E_{v'J'}-E_{v''J''})^3 |M^{\alpha}|^2
,\end{equation}
where the value of $A$ is per second \citep{Abgrall2000}.

For the analysis of quasar data, damping coefficients $\Gamma_i$ are important, corresponding to the total radiative decay rates of the excited states which are computed from the sum of the Einstein coefficients emitted from a specific excited level  ($v',J'$). This total radiative decay rate:
\begin{equation}
  \Gamma_i^{v'J'} = \Sigma_{v''J''} A_{v'J',v''J''} + \Sigma_{J''} \int A_{v'J'}(E)dE
,\end{equation}
includes both bound-bound transitions as well as bound-free transitions, where the integration over the kinetic energy $E$ spanning the continuum of the $X^1\Sigma_g^+$ ground electronic state is performed from $E=0$, corresponding to the dissociation limit of H$_2$.
These different parameters have been checked against experimental studies of H$_2$ UV emission following electronic excitation \citep{Liu2000,Jonin2000}.

A fourth ingredient for analyzing H$_2$ spectra for a possible variation of the proton-to-electron mass ratio $\mu$ is the sensitivity coefficient, which for each line is defined as:
\begin{equation}
  K_i = \frac{d \ln \lambda_i}{d \ln \mu}.
\end{equation}
After a first calculation of $K_i$ coefficients based on rotational constants \citep{Varshalovich1993}, a more sophisticated model was developed based on Dunham parameters representing the accurately measured energy level structure~\citep{Ubachs2007}. The latter model took into account nonadiabatic couplings in a semi-empirical way.  Comparison with values based on a two-state-coupling \emph{ab initio} calculation \citep{Meshkov2006} produced agreement between methods within $3 \times 10^{-4}$, providing confidence in the reliability of the $K_i$ values determined. Thereafter $K_i$ coefficients were calculated in a four-state-coupling calculation \citep{Abgrall1993c,Abgrall2000}, comprehensive results of which are considered the most accurate and estimated to be accurate to within $\Delta K_i \sim 3 \times 10^{-4}$~\citep{Salumbides2015}. These values are included in the present listing.

The molecular data for H$_2$, for the four relevant parameters $\lambda_i$, $f_i$, $\Gamma_i$ and $K_i$, are collected in the tables provided here. Data are given for all transitions in the Lyman bands (Table 1) and Werner bands (Table 2) originating in the eight lowest rotational quantum states $J=0-7$, for $v=0$. In quasar spectra, typically, lines are observed for quantum states up to $J=5$. The first column in the tables refers to the identification of the line in four units: (i) L or W for Lyman or Werner; (ii) the vibrational quantum number $v'$ of the excited state; (iii) P, Q, or R for the rotational transition; (iv) the $J$ quantum number of the ground level. The molecular data are represented for $\lambda_i$ in units of \AA, the uncertainty $\Delta\lambda$ also in units of \AA, $f_i$ as a dimensionless line strength, $\Gamma_i$ in units of s$^{-1}$, and $K_i$ as dimensionless numbers. The spectral information is limited to absorption lines that fall redward of the Lyman cutoff, at $\lambda > 912$ \AA. The Lyman bands are known to exhibit only R- and P-branch lines, while the Werner bands exhibit also Q-lines, besides the R- and P-branch lines.

\section{Molecular data for HD}

Besides H$_2$ lines HD transitions are also observed in the line-of-sight of quasars. However, HD is present at low abundance in high-redshift galaxies, well below earth abundances and mostly below galactic abundances~\citep{Tumlinson2010}. In addition, in view of the small dipole moment of HD the state populations equilibrate to the local temperatures, to the effect that in virtually all observations only R(0) lines are observed, probing the $J=0$ population. Only in a single case has the R(1) line of HD been observed~\citep{Balashev2010}. Therefore, only information on R(0), R(1) and P(1) lines and, for the Werner bands also the Q(1) line, is compiled for HD in Table 3.

Values for the transition wavelengths $\lambda_i$ are obtained from vacuum ultraviolet laser spectroscopy~\citep{Ivanov2008}. That study encompasses lines in the Lyman bands for vibrations $v'=0-9$ and for $v'=16$ and, for the Werner band, only $v'=0$, all claimed to be accurate to $\Delta\lambda/\lambda < 5 \times 10^{-8}$. It is noted that the assignments of P and R lines in the $B-X$ ($16,0$) band were erroneously interchanged. These most accurate entries for HD wavelengths are labeled with "1" in Table 3.

For the remaining lines not covered in the laser-based study~\citep{Ivanov2008}, there are two sets with data of equal accuracy, claimed at $\Delta\lambda/\lambda < 4 \times 10^{-7}$. The laser-based study by \cite{Hinnen1995} reports lines for the Lyman bands up to $v'=19$ and Werner bands for $v'=2$ and $v'=4$. These entries are labeled with "2" in Table 3. The remaining lines redward of the Lyman cutoff (up to $v'=21$ for the Lyman bands and up to $v'=5$ for the Werner bands) were taken from a study employing a vacuum ultraviolet Fourier-transform spectrometer fed by synchrotron radiation~\citep{Ivanov2010}. Some values of lower accuracy in the laser study \citep{Hinnen1995} were replaced by those from the synchrotron study, where the latter were found to be more accurate \citep{Ivanov2010}. The entries from the synchrotron study are labeled with "3" in Table 3.
It is noted further that the absorption spectrum of HD exhibits some extra lines due to symmetry breaking in the hetero-nuclear hydrogen species, in excitation to the $EF^1\Sigma_{(g)}^+$ state. Such weak lines were reported in \cite{Dabrowski1976} and \cite{Hinnen1995}, but have not been observed in quasar absorption spectra.

Two independent four-state nonadiabatic perturbation calculations were performed. The first focused on calculations of $f_i$ and Einstein coefficients $A_{v',v"}$, also yielding values for the damping parameters $\Gamma_i$ \citep{Abgrall2006}. The second focused on determining values of $K_i$ coefficients \citep{Ivanov2010}. Values from both studies are adopted in Table 3.

\section{Tables}

The molecular data are presented in three tables:

      - Table 1: Line list and molecular data for the $B^1\Sigma_u^+$ - $X^1\Sigma_g^+$ Lyman bands of H$_2$.

      - Table 2: Line list and molecular data for the  $C^1\Pi_u$ - $X^1\Sigma_g^+$ Werner bands of H$_2$.

      - Table 3: Line list and molecular data for the HD molecule.
The molecular data in these tables supersede previous compilations with updated values, such as the often-used listing by \cite{Malec2010}. Some previous data were not the most accurate, like for transitions with high-$J$ in H$_2$, or were erroneous, like the $f_i$ values for the P- and R branches of the H$_2$ Werner bands. Most of the latter lines had $f_i$ values off by some 10\% or less, except for the R(0) lines, which were off by up to a factor of two. The information on HD is largely updated, since information on many lines not observed yet in quasar absorption studies had not been presented.

\section{Implementation in analysis of quasar absorption spectra}

A search for a possible variation of the proton-to-electron mass ratio can be made operational by comparing fitted values of the line centers $\lambda_i^z$ in absorption spectra of molecular hydrogen in galaxies in the line-of-sight toward quasars with laboratory values $\lambda_i^0$ via:
\begin{equation}
  \frac {\lambda_i^z}{\lambda_i^0} = (1 +z_{abs})\left(1+ \frac{\Delta\mu}{\mu}K_i\right)
,\end{equation}
where $z_{abs}$ is the redshift of the absorbing cloud. In fitting routines, for each specific absorption line the value of $K_i$ is included to derive a constraint on $\Delta\mu/\mu$.

A fitting procedure may build a synthetic spectrum from the molecular data, including wavelengths $\lambda_i$, line strengths $f_i$,  and sensitivity coefficients $K_i$. In the model spectrum, a value for a Doppler broadening parameter $b$ is built in, convolved with the instrument width and the Lorentzian component $\Gamma_i$, giving each line a composite Voigt profile. This parameter $\Gamma_i$ impacts in cases of strong saturated absorption where the absorption line exhibits the typical damping profile. For each ground-state level a population parameter $N(J)$ is attached, which connects the information of multiple lines probing the same level in a physical way. Such a synthetic spectrum is assigned a redshift parameter $z_{abs}$. This method, referred to as "comprehensive fitting method", is coded and implemented in a dedicated open-source fitting suite VPFIT developed by \citet{Carswell2014}. It allows for inclusion of more than one velocity component, each component assigned with physical parameters, such as $b'$ and $N'(J)$, as well as a redshift parameter $z_{abs}'$. Asymmetric line shapes due to overlapping velocity components can therefore be treated. In addition, the spectral contribution of neutral H I lines, as well as metal lines associated with Fe, Cr, and so on can be included in the fitting routine.

In order to verify the effect of the updates and corrections presented in the molecular data files of Tables 1, 2, and 3, a re-analysis of two quasar absorption spectra was performed.
The quasar absorption spectrum toward J1237+0647 was chosen because it contains two R(0) and two R(1) Werner lines of H$_2$, as well as two R(0) Werner lines of HD; the intensities of these lines are most strongly affected by the molecular data update. J1237+0647  was recently analyzed by \cite{Dapra2015} to yield a constraint of $\Delta\mu/\mu = (-5.54 \pm 6.30_{stat}) \times 10^{-6}$ using the older molecular data set \citep{Malec2010}. We note that the uncertainties are expressed in terms of $1\sigma$ standard deviations.
Again, VPFIT was used with exactly the same settings, and the same treatment of the spectrum, with the same H I assignments, metal absorbers, and velocity components. The re-fit yields a value of $\Delta\mu/\mu = (-4.37 \pm 6.30_{stat}) \times 10^{-6}$, hence a difference of the value of $\Delta\mu/\mu$ by less than 20\% of the standard deviation, while the reduced $\chi^2_{\nu}$ was decreased from $1.327$ to $1.318$. The latter indicates a slight improvement of the fit as a result of the renewed molecular data set.

In addition, the quasar absorption spectrum towards B0642-5038, previously analyzed by \cite{Bagdonaite2014}, was refitted. This spectrum contains only a single R(0) Werner line affected by an intensity change. A specific analysis was performed isolating low- and high-$J$ transitions for a one-velocity-component model. For the original molecular data set \citep{Malec2010}, a value of $\Delta\mu/\mu = (+9.88 \pm 6.14_{stat}) \times 10^{-6}$ was obtained, and for the new molecular data set this value was found to be $\Delta\mu/\mu = (+9.44 \pm 6.09_{stat}) \times 10^{-6}$ . The difference between the results amounts to 7\% of a standard deviation with the reduced $\chi^2_{\nu}$ marginally decreasing from $1.189$ to $1.188$. The fact that the re-analysis of B0642-5038 produces a smaller change than that of J1237+0647 may be attributed to the lesser amount of affected R(0)-R(1) Werner lines in the spectrum.

Overall, these results demonstrate that the correction to the molecular data set has no significant impact on the previous constraints on a varying proton-to-electron mass ratio and the conclusions drawn from previous analyses as discussed in \cite{Ubachs2016}.
Apart from the specific application to quasar absorption spectra, the updated and accurate molecular data for H$_2$ and HD may also be applied to investigations of gamma-ray burst afterglows~\citep{Bolmer2018}.

\section{Conclusion}

A molecular data set is compiled for the analysis of H$_2$ and HD absorption spectra in the Lyman and Werner bands redward of the Lyman cutoff. The most accurate values to date are compiled, for transition wavelengths $\lambda_i$, line oscillator strengths $f_i$, and radiative damping coefficients $\Gamma_i$ for all spectral lines, as well as sensitivity coefficients $K_i$ to a variation of the proton-to-electron mass ratio $\mu$.
The tables provide the required zero-redshift information for the analysis of quasar absorption spectra to retrieve a constraint on $\Delta\mu/\mu$ on a cosmological timescale.

\begin{acknowledgements}
The authors wish to thank S. A. Balashev for bringing errors and inconsistencies in the previous molecular data list to our attention. J. Bagdonaite and M. Dapr\`{a} are thanked for discussions and clarifications.
W.U. received funding from the European Research Council (ERC) under the European Union's Horizon 2020 research and innovation program (Grant No. 670168).
\end{acknowledgements}

% WARNING
%-------------------------------------------------------------------
% Please note that we have included the references to the file aa.dem in
% order to compile it, but we ask you to:
%
% - use BibTeX with the regular commands:
%   \bibliographystyle{aa} % style aa.bst
%   \bibliography{Yourfile} % your references Yourfile.bib
%
% - join the .bib files when you upload your source files
%-------------------------------------------------------------------

\end{document}